\documentclass[%
 reprint,
 amsmath,amssymb,
 aps,
]{revtex4-2}

\usepackage{layouts}
\usepackage{natbib}
\setcitestyle{authoryear,open={(},close={)}}
\usepackage[dvipsnames]{xcolor}
\definecolor{mycitecolor}{HTML}{15998E}
\usepackage[colorlinks=true,linkcolor=blue,citecolor=mycitecolor]{hyperref}%

\usepackage{graphicx}% Include figure files
\usepackage{dcolumn}% Align table columns on decimal point
\usepackage{bm}% bold math
\usepackage{afterpage}
\newcommand\myemptypage{
    \null
    \thispagestyle{empty}
    \addtocounter{page}{-1}
    \newpage
    }
    
\providecommand{\keywords}[1]
{
  \small	
  \textbf{\textit{Keywords---}} #1
}

\newcommand{\dat}{\textbf{d}}
\newcommand{\parvec}{\boldsymbol{\theta}}
\newcommand{\xvec}{\textbf{{x}}}
\newcommand{\muvec}{\boldsymbol{\mu}}
\newcommand{\like}{\mathcal{L}}
\newcommand{\Cinv}{\textbf{C}^{-1}}
\newcommand{\Covar}{\textbf{C}}
\newcommand{\deltdat}{\boldsymbol{\Delta}}
\newcommand{\fisher}{\textbf{F}}
\DeclareMathOperator{\tr}{tr}

\begin{document}

\preprint{APS/123-QED}

\title{Lossless, Scalable Implicit Likelihood Inference for Cosmological Fields}% Force line breaks with \\

\author{T. Lucas Makinen}
\email{timothy.makinen@cfa.harvard.edu}

\affiliation{%
 Sorbonne Université, CNRS, UMR 7095, Institut d’Astrophysique de Paris, 98 bis boulevard Arago, 75014 Paris, France
}%

\affiliation{Harvard \& Smithsonian Center for Astrophysics,
Observatory Building E, 60 Garden St, Cambridge, MA 02138, United States}

\author{Tom Charnock}%
\affiliation{%
 Sorbonne Université, CNRS, UMR 7095, Institut d’Astrophysique de Paris, 98 bis boulevard Arago, 75014 Paris, France
}%

\author{Justin Alsing}%
\affiliation{%
Oskar Klein Centre for Cosmoparticle Physics, Department of Physics, Stockholm University, Stockholm SE-106 91, Sweden
}
\affiliation{
Imperial Centre for Inference and Cosmology (ICIC) \& Astrophysics Group,
Imperial College London, Blackett Laboratory, Prince Consort Road, London SW7 2AZ,
United Kingdom
}

\author{Benjamin D. Wandelt}
\email{bwandelt@iap.fr}
\affiliation{%
 Sorbonne Université, CNRS, UMR 7095, Institut d’Astrophysique de Paris, 98 bis boulevard Arago, 75014 Paris, France
}%
\affiliation{%
Center for Computational Astrophysics, Flatiron Institute, 162 5th Avenue, New York, NY 10010, USA
}%

\date{\today}% It is always \today, today,

\begin{abstract}
We present a comparison of simulation-based inference to full, field-based analytical inference in cosmological data analysis.  To do so, we explore parameter inference for two cases where the information content is calculable analytically:  Gaussian random fields whose covariance depends on parameters through the power spectrum; and correlated lognormal fields with cosmological power spectra. We compare two inference techniques: i) explicit field-level inference using the known likelihood and ii) implicit likelihood inference with maximally informative summary statistics compressed via Information Maximising Neural Networks (IMNNs). We find that a) summaries obtained from convolutional neural network compression do not lose information and therefore saturate the known field information content, both for the Gaussian covariance and the lognormal cases, b) simulation-based inference using these maximally informative nonlinear summaries recovers nearly losslessly the exact posteriors of field-level inference, bypassing the need to evaluate expensive likelihoods or invert covariance matrices, and c) even for this simple example, implicit, simulation-based likelihood incurs a much smaller computational cost than inference with an explicit likelihood. This work uses a new IMNN implementation in \texttt{Jax} that can take advantage of  fully-differentiable simulation and inference pipeline. We also demonstrate that a single retraining of the IMNN summaries effectively achieves the theoretically maximal information, enhancing the robustness to the choice of fiducial model where the IMNN is trained.

\end{abstract}


%Use showkeys class option if keyword
                             % display desired
\maketitle

%\tableofcontents

\section{\label{sec:intro}Introduction}

Modern astronomical and cosmological surveys consist of enormous raw dataset too large to be
interrogated directly. Most of the time, cosmological analyses require that raw data be compressed to
a computationally tractable set of summaries \citep{ Tegmark_1997,alsing2018_general}. Physical understanding can often guide the choice of this compression, such as light curve compression
and power spectrum computation. However, for data for which the physical processes are poorly
understood, choosing an optimal compression that preserves the maximal amount of information from the
raw data is tricky. Existing techniques such as Principal Component Analysis \citep{Connolly1994ph, francis_pca} and power spectrum
estimation have proven successful for astrophysical problems, but due to the sheer size of some
datasets, even summary sets prove too large to work with in the context of likelihood computation.
Surveys such as Euclid are forecast to produce $10^4$ summary statistics \citep{massive_Heavens_2017}. Upcoming astrophysical surveys such as Euclid \citep{laureijs2011euclid}, the Legacy Survey of Space and time (LSST) \citep[]{lsstdarkenergysciencecollaboration2012large} and the Square Kilometre Array \citep[]{ska_Weltman_2020} will allow the
cosmological field to be interrogated directly in the context of dark energy measurement and
large-scale structure formation. For these non-Gaussian fields the power spectrum alone is not
sufficient to describe galaxy clustering, posing a challenge for cosmologists. Recent advances in deep learning have made interpreting large swaths of cosmological data more tractable, from emulators \citep{shirleyd3m, Kodi_Ramanah_2020, moster2020galaxynet, deoliveira2020fast} to systematics and foreground removal \citep{makinen2020deep21, Puglisi_2020, Petroff_2020} (see \cite{ntampaka2021role} for a recent review).

We study Information Maximising Neural Networks (IMNNs) \citep{Charnock_IMNN} in the context of computing posteriors for cosmological parameters from cosmological fields. IMNNs are neural networks that compress data to informative nonlinear summaries, trained on simulations to maximise the Fisher information. Training such networks automatically gives Gaussian approximation uncertainties from the summary Fisher information, and score estimates for parameters can then be used for efficient implicit likelihood inference. Beyond cosmology, IMNNs perform massive dimensionality reduction, even for highly non-Gaussian problems, such as galaxy type identification from multiband images \citep{livet2021catalogfree}.

Other studies have investigated neural techniques for point estimate cosmological parameter extraction from cosmological fields via regression networks trained on simulation-parameter pairs \citep{pan2020cosmological, ravanbakhsh2017estimating, Kwon_2020, prelogovic2021machine, Fluri_2019lensing, Fluri_2018, Matilla_2020, ribli2018improved, Gillet_2019} with squared loss. As reviewed in \cite{villaescusanavarro2020neural}, these techniques can estimate the posterior mean of parameters. This implies they require simulations drawn from a prior, specified at the time of training, not just near the parameters favored by the data. This adds to the variability that needs to be fit by the network. % but do not guarantee an optimal compression of the cosmological field for efficient Bayesian posterior estimation.

We show that IMNNs, trained on simulations at a fiducial point in parameter space, can compress cosmological fields to maximally informative score estimates \citep{alsing2018_general} without specifying a prior at the training stage.  Lossless field-level inference--equivalent to using all pixels in an image--can then be performed without evaluating an explicitly specified (or expensive or intractable) likelihood and while specifying the prior at the following inference stage. We first introduce IMNNs within a differentiable framework when exact simulation gradients are accessible.
We  compare inference with IMNN-compressed parameters to a computationally expensive yet exact full field-level
inference for  Gaussian fields, demonstrating
that a convolutional neural network can be trained to extract the maximal amount of information from Gaussian fields using a modest number of simulations. Although simplistic, Gaussian fields are relevant to early cosmological density fields such as the Cosmic Microwave Background (CMB) as measured by \cite{planck_collab}. We verify through Approximate Bayesian Computation \citep{grazian2020review,Cranmer30055} with these summaries that we recover the exact analytic posterior. We then
demonstrate an extension to log-normal fields generated with a  cosmological power
spectrum.  We demonstrate that IMNN compression extracts the maximal amount of information
from raw
field inputs, and produces Fisher matrices that allow for efficient Density Estimation Likelihood Free Inference (DELFI), recovering simulation-generated posteriors with a moderate number of simulations. Finally, we show that IMNN compression is robust to the choice of the fiducial model - even for poor choices of fiducial model a single re-training iteration on score estimates recovers optimality.

After reviewing and defining additions to our IMNN framework in Section \ref{sec:imnn} and outlining simulation-based inference techniques in Section \ref{sec:sbi} we proceed to applying our framework to progressively more difficult problems to verify our techniques in Sections \ref{sec:benchmark} and conclude in \ref{sec:cosmo}.

\section{Information Maximising Neural Networks}\label{sec:imnn}

\begin{figure*}[ht!]
    \centering
    \includegraphics[width=\textwidth]{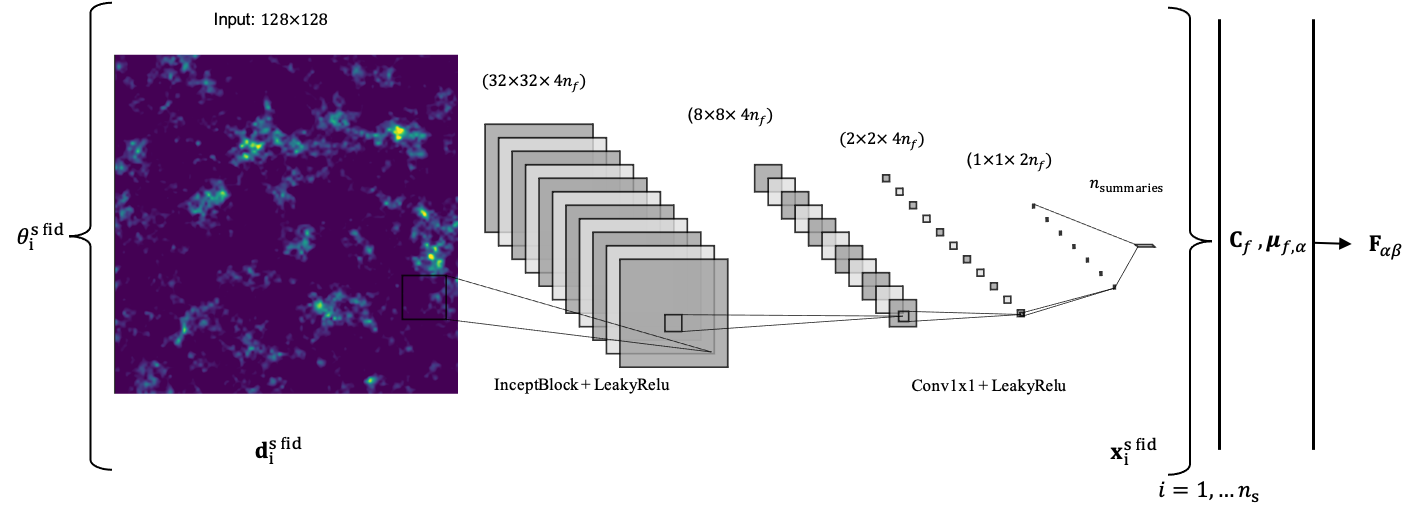}
    \caption{Cartoon of the information maximising neural network scheme. Raw real-space field
    values are fed directly to the network compressor. We compute field simulations $\dat^{\rm s fid}_i$ at
    the fiducial model and pass the data through the network to produce $n_{\rm summaries} = 2$
    summary statistics $\xvec_i^{\rm s fid}$, for each simulation. This is done for $n_s$ independent simulations evaluated at $\parvec_{\rm fid}$. Since both the simulator and compression steps are numerically differentiable, the gradients with respect to input parameters are readily obtained to compute equation \ref{eq:autograd-enabled} which are used to compute the covariance $\textbf{C}_f$ and derivatives of the summaries $\boldsymbol{\mu}_{f,\alpha}$. The output of the IMNN is Fisher information matrix, computed via equation \ref{eq:fisher-compressed}. The network is trained via gradient back-propagation, with the $\det \textbf{F}$ and $\textbf{C}_f$ contributing to the scalar loss function.}
    \label{fig:conv-scheme}
\end{figure*}

The IMNN framework is presented in full in \cite{Charnock_IMNN},
but we review the formalism here for completeness, as well as
introduce two new aspects to the technique. The sharper the peak of
an informative likelihood function $\mathcal{L}(\textbf{d} | 
\parvec )$ for some data $\textbf{d}$ with $n_{\textbf{d}}$ data
points and $n_{\parvec}$ parameters at a given value of $\parvec$,
the better $\parvec$ is known. The Fisher information matrix
describes how much information $\dat$ contains about the
parameters, and is given as the second moment of the score of the
likelihood
\begin{equation}
    \textbf{F}_{\alpha \beta} = \int d \dat\ \mathcal{L}(\dat | \parvec) \frac{\partial \ln \mathcal{L}(\dat | \parvec)}{\partial \theta_\alpha} \frac{\partial \ln \mathcal{L}(\dat | \parvec)}{\partial \theta_\beta},
\end{equation}
and can be written as
\begin{equation}\label{eq:fisher-def}
    \textbf{F}_{\alpha \beta} = -\left\langle \frac{\partial^2 \ln \mathcal{L}}{\partial \theta_\alpha \partial
\theta_\beta} \right\rangle \Big|_{\parvec = \parvec_{\rm fid}} .
\end{equation}
A large Fisher information for a set of data indicates that the data is very informative about the model
parameters attributed to it. Fisher forecasting for a given model is made possible by the Cram\'er-Rao bound \citep{cramerharald_1946, rao_1945}, which states that the minimum variance of the value of an estimator $\parvec$ is given by 
\begin{equation}\label{eq:cramer-rao}
    \langle (\theta_\alpha - \langle \theta_\alpha \rangle ) (\theta_\beta - \langle \theta_\beta
    \rangle) \rangle \geq \textbf{F}^{-1}_{\alpha \beta}.
\end{equation}

We will write the compression as a function $f:\dat \rightarrow \xvec$. For large datasets, data compression is essential for inference to avoid the curse of dimensionality. The MOPED formalism \citep{Heavens_2000} gives optimal score compression for cases where the likelihood is  well-approximated by a Gaussian form. 

IMNNs are neural networks that optimize data compression and compute the Fisher information of a data set even if the data likelihood is unknown or intractable, simply based on having simulations of the data at a given fiducial parameter point and local information about how the parameters change the data distribution. It can be shown (Wandelt, 2021, in preparation) that the optimality of the IMNN summaries holds for any unknown or intractable data likelihood even though the IMNN maximizes Fisher information assuming the Gaussian likelihood form for the IMNN summaries
\begin{equation}\label{eq:og-moped-like}
    -2 \ln \mathcal{L}(\xvec | \dat) = (\xvec - \muvec_f(\parvec))^T \textbf{C}_f^{-1}(\xvec - \muvec_f(\parvec))
\end{equation}
where 
\begin{equation}
    \muvec_f(\parvec) = \frac{1}{n_s} \sum_{i=1}^{n_s} \xvec_i^s
\end{equation}
is the mean of the compressed summaries $\textbf{x}_i^s$, with $\{\textbf{x}^s_i | i \in [1, n_s] \}$. Here $i$ indexes the random initialisation of $n_s$ simulations. The data are obtained via simulation $\dat^s_i = \dat_i^s(\parvec, i)$ via the compression scheme $f:\dat_i^s \rightarrow \textbf{x}_i^s$. The covariance of the summaries is computed from the data as well:
\begin{equation}
    (\textbf{C}_f)_{\alpha \beta} = \frac{1}{1- n_s}\sum_{i=1}^{n_s} (\xvec_i^s - \muvec_f)_\alpha (\xvec_i^s - \muvec_f)_\beta.
\end{equation}
A modified Fisher matrix can then be computed from the likelihood in equation \ref{eq:og-moped-like}:
\begin{equation}\label{eq:fisher-compressed}
    \textbf{F}_{\alpha \beta} =  \tr [\muvec_{f,\alpha}^T C^{-1}_f \muvec_{f, \beta}],
\end{equation}
where we introduce the notation $\bm{y}_{,\alpha}\equiv\frac{\partial \bm{y}}{\partial\theta_\alpha}$ for partial derivatives with respect to parameters. If the compression function $f$ is a neural network parameterized by layer weights $\textbf{w}^\ell$ and biases
$\textbf{b}^\ell$ (with $\ell$ the layer index), the summaries (and respective mean and covariance) then become
functions of these new parameters $\xvec(\parvec) \rightarrow \xvec(\parvec, \textbf{w}^\ell, \textbf{b}^\ell)$. To evaluate equation \ref{eq:fisher-compressed} for a neural compression, we must compute
\begin{equation}
    \muvec_{f,\alpha} = \frac{\partial}{\partial \theta_\alpha} \frac{1}{n_s}\sum^{n_s}_{i=1} \xvec^{s\ \rm fid}_i 
\end{equation}

One way of computing the derivatives of the summary means with respect to the parameters is to define a finite difference gradient dataset by altering simulation fiducial values by a small amount, yielding
\begin{equation}\label{eq:discrete-deriv}
    \left( \frac{\partial \hat\mu_i}{\partial \theta_\alpha} \right)^{s\ \rm fid} \approx \frac{1}{n_s}\sum^{n_s}_{i=1} \frac{\xvec^{s\ \rm fid +}_i - \xvec^{s\ \rm fid -}_i}{\Delta \theta^+_\alpha - \Delta \theta^-_\alpha}.
\end{equation}
However, this discrete numerical differentiation requires more memory ($\theta_\alpha^{\rm fid \pm}$ datasets), as well as the hyperparameter tuning for the size of the numerical derivatives. 

An alternative we will explore in this paper is to calculate the adjoint gradient of the simulations as well as the derivatives of the network parameters with respect to the simulations:
\begin{equation}\label{eq:autograd-enabled}
    \muvec_{f,\alpha} =  \frac{1}{n_s}\sum^{n_s}_{i=1} \left( \frac{\partial \xvec}{\partial \theta_\alpha} \right)_i^{s\ \rm fid} = \frac{1}{n_s} \sum^{n_s}_{i=1} \sum^{n_d}_{k=1} \frac{\partial \xvec^{s\ \rm fid}_i}{\partial d_k}\frac{\partial \dat^{s\ \rm fid}_i}{\partial \theta_\alpha} .
\end{equation}
If the gradient of the simulations can be computed efficiently, this technique for computing the compression Fisher information eliminates the need for hyperparameter tuning of the finite difference derivative size, $\Delta \theta_\alpha$.

The network compression is trained to maximise the determinant of the Fisher information, computed via equation \ref{eq:fisher-compressed}.
As described in \citeauthor{Charnock_IMNN} and \citeauthor{livet2021catalogfree}, the Fisher information is invariant to nonsingular linear transformations of the summaries. To remove this ambiguity, a term driving covariance to the identity matrix is added
\begin{equation}
    \Lambda_C = \frac{1}{2} \left( \left|\left|(\textbf{C}_f-\mathbf{1})\right|\right|^2_\mathcal{F} + \left|\left|(\textbf{C}^{-1}_f-\mathbf{1})\right|\right|^2_\mathcal{F} \right),
\end{equation} 
where  $||\bm{A}||_\mathcal{F}\equiv \sqrt{\tr{\bm{AA}^T}}$ denotes the Frobenius norm. This yields the loss function
\begin{equation}\label{eq:imnn-loss}
\Lambda = -|\det \textbf{F} | + r_{\Lambda_C} \Lambda_C,
\end{equation}
with regularization parameter
\begin{equation}
    r_{\Lambda_C} = \frac{\lambda \Lambda_C}{\Lambda_C + \exp(-\alpha \Lambda_C)},
\end{equation}
where $\lambda$ and $\alpha$ are user-defined parameters. When the covariance is far from identity, the $r_{\Lambda_C}$ function is large and the optimization focuses on bringing the
covariance and its inverse back to identity. The network is trained until the Fisher information stops increasing for a pre-determined number of iterations.

\subsection{On-the-fly training }
For this study we introduce some modifications to the IMNN implementation, including sampling and simulations written in \texttt{Jax}. The \texttt{Jax} XLA backend allows for efficient GPU-enabled numerical differentiation. If simulations are exactly and efficiently differentiable, equation \ref{eq:autograd-enabled} can be used to eliminate the need for the two copies of derivatives needed for $\{ \theta_\alpha^{\rm fid \pm} \}$ in equation \ref{eq:discrete-deriv}. For cheap simulators, \texttt{Jax}'s auto-differentiation allows simulation realizations and their derivatives to be generated anew each epoch for neural network training, additionally eliminating the need for a validation dataset.

\subsection{Iterative IMNN training}\label{subsec:iterative-training}
In cases where the fiducial model is poorly chosen for the compression, e.g. $\parvec_{\rm target}$ is very far from $\parvec_{\rm fid}$, we still wish our IMNN to be able to estimate the correct posterior. In these circumstances, an iterative training scheme can be adopted, namely
\begin{enumerate}
    \item Train IMNN on $\parvec_{\rm fid}$ and perform a Gaussian Approximation on target data using the IMNN's Fisher matrix and score estimation
    \item If $\parvec_{\rm MLE}$ is more than $ 1\sigma$ from $\parvec_{\rm fid}$, retrain the IMNN using $\parvec_{\rm fid} \gets \parvec_{\rm MLE}$. If not, stop.
\end{enumerate}
We explore this method in the context of dark-matter-like field inference in Section \ref{sec:cosmo}.

\section{Simulation-Based Inference Techniques}\label{sec:sbi}
For many astronomical and cosmological problems, it can become too difficult to write down analytic likelihood functions to describe both physics and survey instrumental effects. In these situations, Approximate Bayesian Computation (ABC) can make inference possible when a likelihood is not available \citep{grazian2020review}. ABC is a technique in which parameters are sampled from a prior and used to forward simulate mock data to be compared to the real data, and accepted if the corresponding mock data (or compressed summaries) are ``close enough'' by a pre-defined metric to the target data or summaries. \citep{Cranmer30055}.

However, ABC methods can scale poorly with large prior volumes and high parameter dimensionality, making the forward simulation step expensive for large simulations like N-body grids. Methods like Population Monte Carlo can improve convergence by drawing from a weighted prior each iteration \citep{kitagawa96}, but are still highly simulation-intensive. Density Estimation Likelihood Free Inference (DELFI) \citep{pydelfiAlsing_2019} instead leverages neural density estimators to parameterize the \textit{summary data likelihood}, $p(\xvec | \parvec)$. These neural networks can be trained efficiently on a smaller number of parameters drawn from the prior and their respective compressed forward simulations. Once $p(\xvec | \parvec)$ is learned, the likelihood can be evaluated at $\xvec_{\rm target}$ and multiplied by the prior to yield the posterior $p(\parvec | \xvec) = p(\xvec | \parvec)\times p(\parvec)$. In this work we primarily consider Conditional Masked Autoregressive Flows (CMAFs). CMAFs are stacks of neural autoencoders carefully masked to parameterize the summary-parameter likelihood. This can be seen by noting that any probability density can be factored as a product of one-dimensional conditional distributions via the chain rule of probability:
\begin{equation}
    p(\textbf{x} | \boldsymbol{\theta}) = \prod_{i=1}^{\dim(\textbf{x})} p({\rm x}_i | \textbf{x}_{1:i-1}, \boldsymbol{\theta}).
\end{equation}
Masked Autoencoders for Density Estimation (MADE) \citep{papa_mafs2017, papamakarios2019sequential} model each of these one-dimensional conditionals as Gaussians with mean and variance parameters parameterized by neural network weights, $\textbf{w}$. The neural network layers are masked in such a way that the autoregressive property is preserved, e.g. the output nodes for the density $p({\rm x}_i | \textbf{x}_{1:i-1}, \boldsymbol{\theta})$ only depend on $\textbf{x}_{1:i-1}$ and $\boldsymbol{\theta}$, satisfying the chain rule. MADEs can then be stacked into CMAFs for stable summary likelihood estimation \citep{pydelfiAlsing_2019}. CMAFs are trained to minimize the log-probability, $-\ln U = -\sum_i \ln p(\xvec_i | \parvec_i; \textbf{w})$, where $i$ indexes over a training batch of $\{\xvec, \parvec \}_i$ pairs.

\section{Benchmarking Inference with 2D Gaussian Fields}\label{sec:benchmark}
\begin{figure}[h!]
    \centering
    \includegraphics[width=0.4\textwidth]{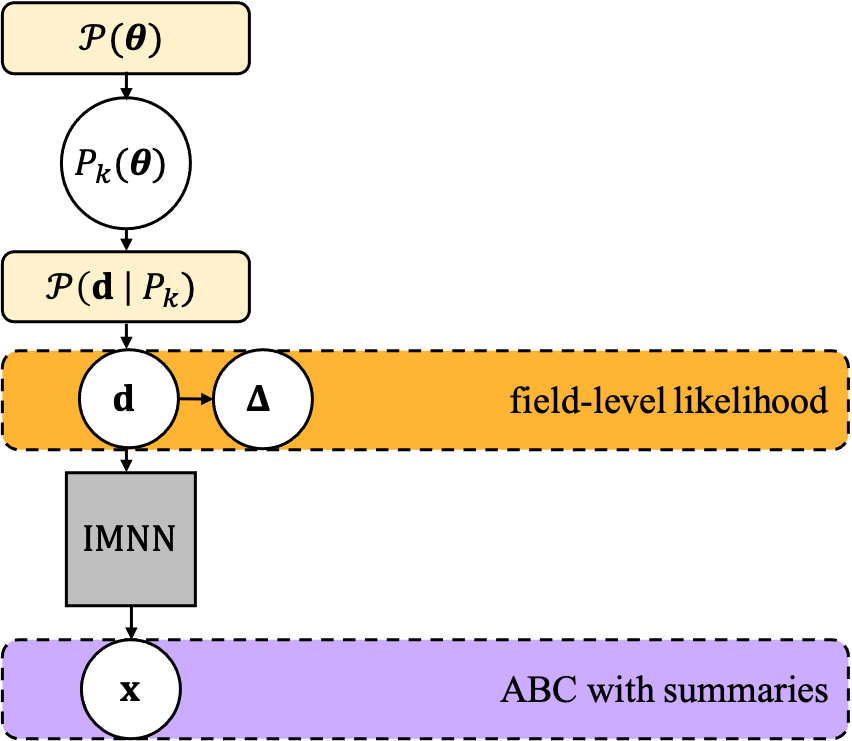}
    \caption{Diagram of the Bayesian hierarchical model and inference techniques for inferring parameters from Gaussian fields. Yellow shaded boxes indicate probability distributions, while arrows denote deterministic connections. The inference schemes considered in this work are denoted by dashed boxes and correspond to their respective contours in figure \ref{fig:128x128-inference}.}
    \label{fig:bhm-gauss}
\end{figure}
We begin verifying a convolutional neural network (CNN) compression scheme within the IMNN framework on 2D Gaussian fields generated from the toy power spectrum model
\begin{equation}\label{eq:toy-powerspec}
    P (k) = Ak^{-B}
\end{equation}
where $\parvec = \{A,B \}$ are the two ``cosmological'' parameters of interest in the field measurement. The parameter $A$ controls the amplitude of the power in Fourier space, while $B$ describes the scale-dependency of the power modes. We compare the implicit likelihood inference using the IMNN compression to a full field-level data assimilation (DA) inference, in which all measured field values are used in the inference to hand.

\subsection{Field-Level Inference via Data Assimilation}
Field-level inference aims to obtain posterior distributions for $A$ and $B$ using all field values $\dat$, indicated by the orange box in the hierarchical model in figure \ref{fig:bhm-gauss}. The likelihood for an $N_{\rm pix}\times N_{\rm pix}$ Gaussian field can be explicitly written down for the Fourier transformed data, $\deltdat$ as: 
\begin{equation}\label{eq:field-like}
    \like{}(\deltdat | \parvec) = \frac{1}{ \det(2\pi\Covar)^{1/2}}\exp{\left(-\frac{1}{2} \deltdat^T \Cinv \deltdat \right)}
\end{equation} where the covariance matrix $\Covar = P(k)$ is just the target power spectrum. 
It is helpful to picture ``unwinding'' the gridded data as
a vector of pixel values of shape $\deltdat \in
\mathbb{R}^{N_p}$, where $N_p = N^2$. At this point our
likelihood is a function of the observed data and the
parameters $\parvec$ governing our theoretical model. Since the entire field is used for the parameter estimation with no intermediate compression step,
this can be viewed as the exact solution to the inference problem.

\subsubsection{Analytic Fisher Matrix}
To test our compression scheme, it is also of interest to obtain the exact information content of the Gaussian Field. For the DA likelihood of equation \ref{eq:field-like}, the Fisher Matrix for parameters indexed by $\alpha, \beta$ is given by \citep{Dodelson_2003}:
\begin{equation}\label{eq:fisher}
    \fisher_{\alpha \beta} = \langle \mathcal{F} \rangle= \frac{1}{2} \tr (\Covar_{, \alpha} \Cinv \Covar_{, \beta} \Cinv),
\end{equation}
where $\textbf{C}$ is the full-rank covariance matrix appearing in equation \ref{eq:field-like} (see Appendix \ref{app:fisher-field-comp} for a detailed calculation). 

For our simple power law case, the covariance between modes is given as:
\begin{equation}
    \textbf{C}_{k_i, k_j} = \left( P (k_i) \right) \delta_{ij} .
\end{equation}
To compute the Fisher information for the two-parameter model $\parvec = (A,B)$, only the first derivatives of the covariance,
\begin{align}
    \frac{\partial P}{\partial A} &= \left( k_i^{-B} \right)\delta_{ij} \\
    \frac{\partial P}{\partial B} &= \left( -Ak_i^{-B}\ln k_i \right) \delta_{ij}
\end{align}
are needed. Evaluating Eq. \ref{eq:fisher} numerically for an $N_{\rm pix}=128$ field in grid units with a $\textbf{k}\in [0.5 / 128, 1.0]$ and $\parvec_{\rm fid} = \{1.0, 0.5\}$ yields a Fisher Information matrix of
\begin{equation}
    \textbf{F} = \begin{pmatrix}
     2047.5  & -1556.2 \\
    -1556.2  & 1743.1
    \end{pmatrix},
\end{equation}
which yields a determinant of 1147281, or a Shannon entropy $\frac{1}{2}\ln \det \textbf{F}= 6.9765$ nats, for $\theta_{\rm fid} = \{1.0, 0.5\}$
We note that we compute this value for all unique modes of the Fourier-transformed field excluding the first (zero) mode.

\subsection{Approximate Bayesian Computation with Optimal Compression}

To test the IMNN compression framework, we also wish to infer $A$ and $B$ via simulation
based-inference with a compression step. The ABC scheme is illustrated in figure \ref{fig:bhm-gauss} via the orange box, with the IMNN compression step denoted by the grey box, yielding summaries $\textbf{x}$. Once an IMNN is trained on a fiducial model, the network
mapping is used to compress raw field data to maximally informative summaries that can be used for learning the likelihood, as in DELFI, or performing
Approximate Bayesian Computation (ABC). Given target data $\dat$ and its corresponding summaries \textbf{\xvec}, parameters are drawn from the priors, in our case the uniform $\mathcal{P}(A)$ and $\mathcal{P}(B)$, used to generate field simulations, which are then fed directly into the network to produce corresponding summaries. The optimal distance measure $\varrho_k^t$ obtained in \cite{alsing2018_general} can be defined using attributes of the IMNN:
\begin{equation}\label{eq:abc-distance}
    \varrho_k^t = \sqrt{(\textbf{x}^{s,t}_{ik}-\textbf{x})^T \textbf{F}_{\rm IMNN} (\textbf{x}^{s,t}_{ik} - \textbf{x})},
\end{equation} where $s$ indexes the summaries, and $i$ labels the random initialisation. Parameter draws are accepted or rejection if the distance $\varrho_k^t$ between simulation and target summaries is within some threshold, $\epsilon$.

\begin{figure}[h!]
    \centering
    \includegraphics[width=0.35\textwidth]{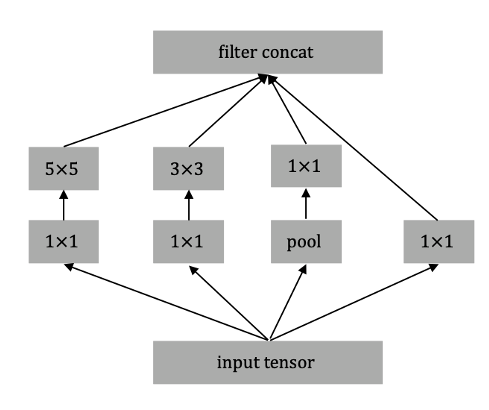}
    \caption{Inception block architecture. We adopt stride-4 downsampling before filter concatenation to quarter the spatial dimensions of the simulations.}
    \label{fig:incept-block}
\end{figure}

\subsubsection{Neural Network Architecture}
With the analytic Fisher Information known, we can test a neural compression scheme within the IMNN framework for the $N_{\rm pix}^2$ field dataset. The network takes as input an $N_{\rm pix}\times N_{\rm pix}$ field, and extracts information via a convolutional compression scheme. We tested several architectures for our chosen $N=128$ simulations, but found an inception-block CNN to be most efficient in training \citep[e.g][]{szegedy2016inceptionv4}. For each inception block, data is passed through $5^2$, $3^2$, $1^2$ convolutions and a \texttt{maxpool} layer in parallel, with outputs concatenated and passed to the next block, shown graphically in figure \ref{fig:incept-block}. We adopt three stride-4 inception blocks, each with $n_f=55$ filters, quartering the output spatial dimensions, each followed by a Leaky \texttt{ReLU} nonlinear activation function. Once spatial dimensions are of shape $(2,2)$, we adopt a stride-2 inception block with $1^2$ kernels followed by a $1^2$ convolution to $n_{\rm summaries} = n_{\rm params} = 2$ summaries.  

To train the network, we generate 200 new simulation realizations on-the-fly each epoch. We make use of \texttt{Jax}'s autograd feature to implement equation \ref{eq:autograd-enabled}, obtaining both simulations and exact derivatives with respect to simulation parameters each epoch for the Fisher information computation. We train the network with the \texttt{Adam} optimizer \citep{kingma_adam_2014} and coupling parameters $\lambda=10$ and $\alpha=0.95$ until the Fisher information stopped increasing for 500 epochs.

\subsection{Numerical Results}
\begin{figure}[htp!]
    \centering
    \includegraphics{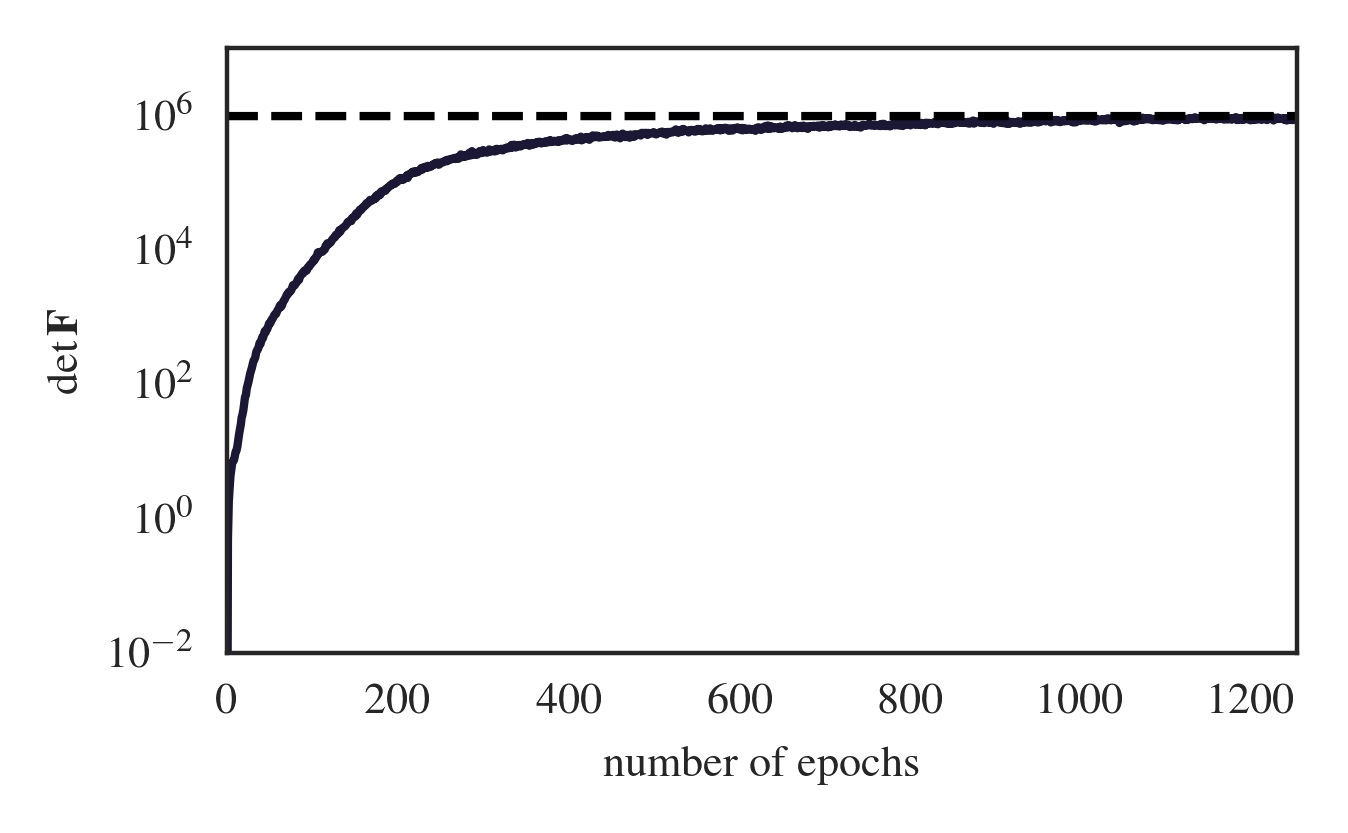}
    \caption{On-the-fly IMNN training using $128\times128$ field simulations.  Within 550 epochs, the network is able to extract 90 \% of the theoretical Fisher information (dashed black line), saturating to $98\%$ by 1500 training epochs.}
    \label{fig:training}
\end{figure}

In this section we present the numerical results obtained from the two inference methods.

\begin{figure}[htp!]
    \centering
    \includegraphics{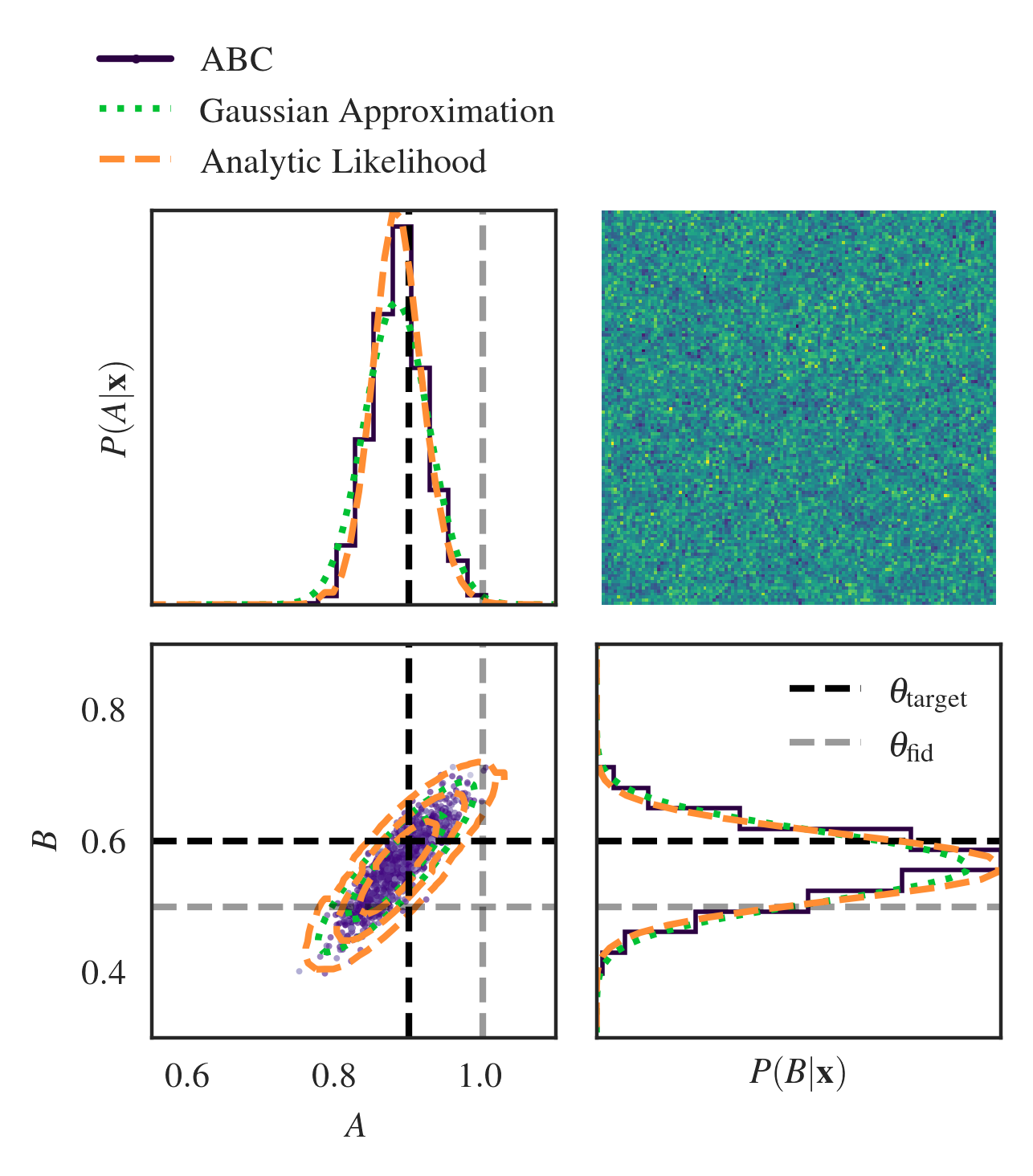}
    \caption{Inference for power spectrum-generated $128\times128$ Gaussian random field. The target
    field ({top right panel}) was generated using a $\parvec_{\rm target} = (0.9,0.6)$, shown
    via the dashed black lines. The Gaussian approximation ({green contour}) and ABC 
    computation ({purple histogram}) were made using an IMNN-calculated Fisher matrix, trained on a
    fiducial
    $\parvec_{\rm fid}=(1.0, 0.5)$, ({dashed grey crosshairs}). Accepted ABC simulations are displayed in the scatterplot, coloured by ABC distance. The ABC computation overlaps
    with the full field likelihood along the entire prior range, indicating that the convolutional
    network has effectively summarised the field data for exact inference.}
    \label{fig:128x128-inference}
\end{figure}

We begin by analysing a single Gaussian field realization generated by $\theta_{\rm target} = \{0.8,
0.8 \} $, shown in the top right panel of figure \ref{fig:128x128-inference}. We first compute the
field-level likelihood from equation \ref{eq:field-like} over the prior range $\mathcal{P}(A) = \mathcal{U}(0.1,
2.0)$ and $\mathcal{P}(B) = \mathcal{U}(0.1, 2.0)$, displayed by the orange contours in figure
\ref{fig:128x128-inference}. We then trained an IMNN from simulations generated from the fiducial model $\parvec_{\rm fid} = \{1.0, 0.5 \} $ until the extracted $\det \textbf{F}$ stopped increasing for 500 epochs (matching the theoretical saturation limit in figure \ref{fig:training}).  Using the trained IMNN, we then use the compressed summaries' Fisher matrix to compute first a Gaussian approximation to $\parvec_{\rm target}$, shown by the green contours in the figure. We then use the IMNN Fisher as a proposal distribution for constructing the simulation-based inference contours via ABC and equation \ref{eq:abc-distance} with $\varepsilon = 0.05$, with accepted points histogrammed in purple. We run 20,000 simulations until a cumulative 1279 proposal parameters were accepted, which takes approximately 20 minutes on a single GPU. 

The IMNN posterior is almost as tight as the field-level inference, indicating that the nonlinear
compression has successfully summarized the field data at the same precision level as using all of the data to hand. 

For this example we verified our technique with Approximate Bayesian Computation, but for the next example we find a massive reduction in the number of simulations needed by employing DELFI techniques for inference. 

\section{Cosmological Parameters from Mock Dark Matter Fields}\label{sec:cosmo}
Here we consider an application of the differentiable IMNN framework making use of a more cosmologically-relevant power spectrum in the context of inferring parameters from a dark matter field. Log-normal fields are often used as an approximation to late-time cosmological overdensity fields \citep[][]{ln-coles-jones, leclercq2021accuracy}. Specifically, we simulate log-normal fields from Gaussian noise \textit{to have a specified power spectrum}, $P_{\rm LN}(k)$ (see Appendix \ref{sec:logfromgauss}). Since the lognormal field is a bijective mapping of the Gaussian field it must conserve the information it contains.  Therefore we can  calculate the theoretical Fisher information content contained in the lognormal field by  equation \ref{eq:fisher} with the covariance of the underlying Gaussian field.  To do so we need to obtain the underlying Gaussian field's power spectrum, $P_{\rm G}$, since this spectrum is diagonal in $k$-space. We employ the backwards conversion formula, presented in \cite{Greiner_2015} to obtain
\begin{equation}\label{eq:backwards-conv}
    P_{\rm G} = \int d^u x e^{i \textbf{k} \cdot \textbf{x}} \ln \left( \int \frac{d^u q}{(2\pi)^u} e^{i \textbf{q} \cdot \textbf{x}} P_{\rm LN}(\textbf{q}) \right),
\end{equation}
where $u=N_{\rm pix}$ is the dimensionality of the space. Once calculated, we can proceed to take derivatives of $P_{\rm G}$ to evaluate equation \ref{eq:fisher} as before.

We simulate scaled log-normal fields with the Eisenstein-Hu linear matter power spectrum \citep[]{Eisenstein_1999} as our $P_{\rm LN}$, implemented differentiably in the \texttt{jax-cosmo} module \citep{jax_cosmo_2020}. We seek to infer the critical matter density parameter, $\Omega_c$, as well as $\sigma_8$, the r.m.s. fluctuation of density perturbations at the 8 $\rm h^{-1} Mpc$ scale. Implementing the simulator and power spectrum in \texttt{Jax} means exact numerical derivatives can be computed for $P_{{\rm G},\Omega_c }$ and $P_{{\rm G},\sigma_8 }$. Evaluating equation \ref{eq:fisher} with these expressions at the fiducial model $(\Omega_c, \sigma_8) = (0.6, 0.6)$ yields a Shannon information of 6.2056. 

We generate four target fields each with cosmological volume $V = L^2 = (250 \ \rm h^{-1}\ Mpc)^2$ on a $128^2$ grid using Planck 2015 parameters \citep[][]{planck2015} at redshift $z=0$ (present day). For this example, we explore the iterative
training method introduced in Section
\ref{subsec:iterative-training}. We adopt an IMNN
pipeline with a neural network identical to the
InceptNet used in the Gaussian Fields example, trained on-the-fly with $n_s = n_d = 200$ at a deliberately far-away fiducial model, $\parvec_{\rm fid, 1} = (0.6, 0.6)$. We train the IMNN until the Fisher information from compressed summaries stops increasing for 500 epochs, and then proceed to making score estimates on the target fields, adopting a new fiducial model, $\parvec_{\rm fid, 2}$ at the mean of the four target estimates. We find that in both training cases the network saturates to above $90\%$ of the respective theoretical field information content.

\subsection{DELFI Setup}
\begin{figure}[h!]
    \centering
    \includegraphics[width=0.39\textwidth]{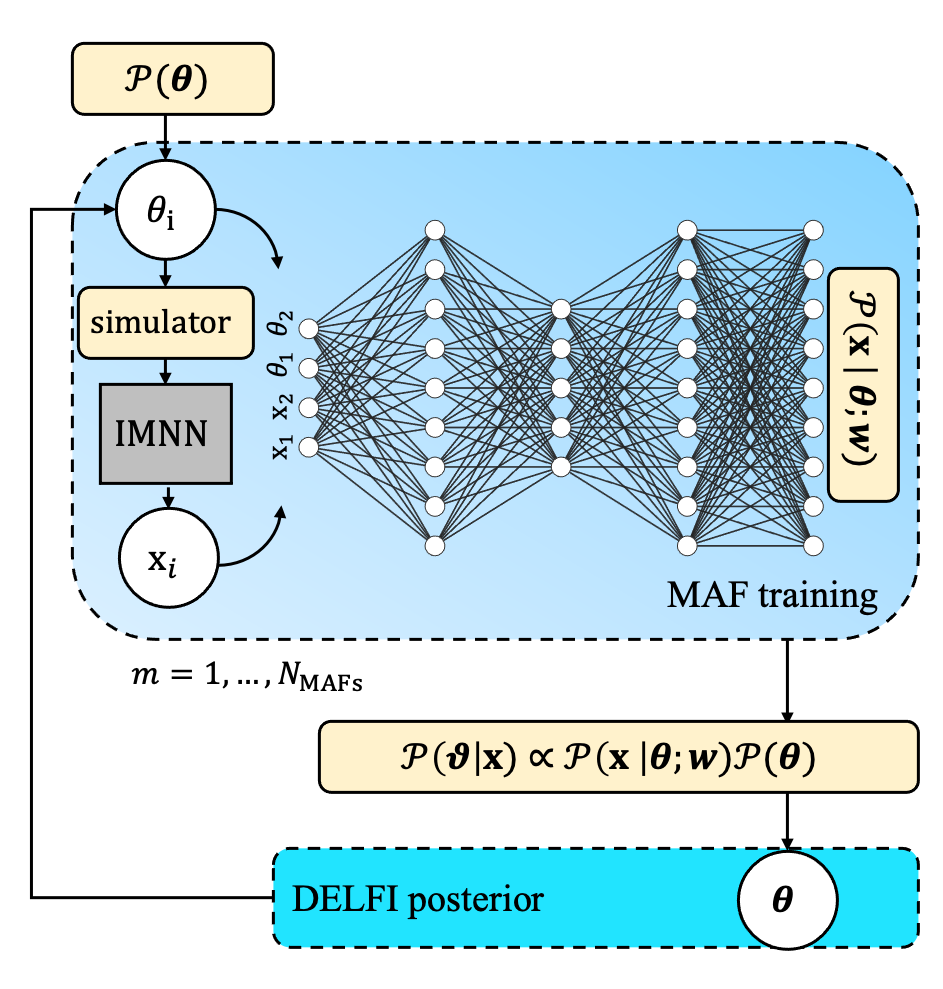}
    \caption{DELFI training setup with IMNN summaries. We train $N_{\rm MAF}=4$ MAFs per target data, first on 1000 simulations from parameters drawn from the prior, then on 3 subsequent subsequent draws from the posterior to produce the blue contours in figure \ref{fig:cosmo-inference}.}
    \label{fig:delfi-setup}
\end{figure}
For this inference we construct Conditional Masked Autoregressive Flows (CMAFs) \citep{papa_mafs2017, papamakarios2019sequential} to perform DELFI posterior estimation for the cosmological parameters, shown graphically in figure \ref{fig:delfi-setup}. For each target field, we utilize an ensemble of 4 networks \citep{deep_ens, pydelfiAlsing_2019} consisting of 5 MADEs, each with two hidden layers of 50 neurons. For our training scheme we first sample 1000 simulations from the prior, simulate fields, compress with an IMNN, and train each DELFI ensemble on batches of 100 $\{\xvec, \parvec \}$ pairs (with a 75-25\% train-validation split) for 1000 epochs with an \texttt{Adam} optimizer with learning rate $10^{-3}$. We then sample the posterior following \citep{papamakarios2019sequential} for new parameters using a  Tensorflow affine-invariant MCMC sampler (adapted from \citep{Foreman_Mackey_2013} and implemented in the \texttt{pydelfi} package \citep{pydelfiAlsing_2019}). Then we simulate  fields from these parameters and compute their compressed summaries. We append the new training data and re-train the CMAFs iteratively until $-\ln U$ stops decreasing significantly.

\subsection{Results}

\begin{figure}[htp!]
    \centering
    \includegraphics{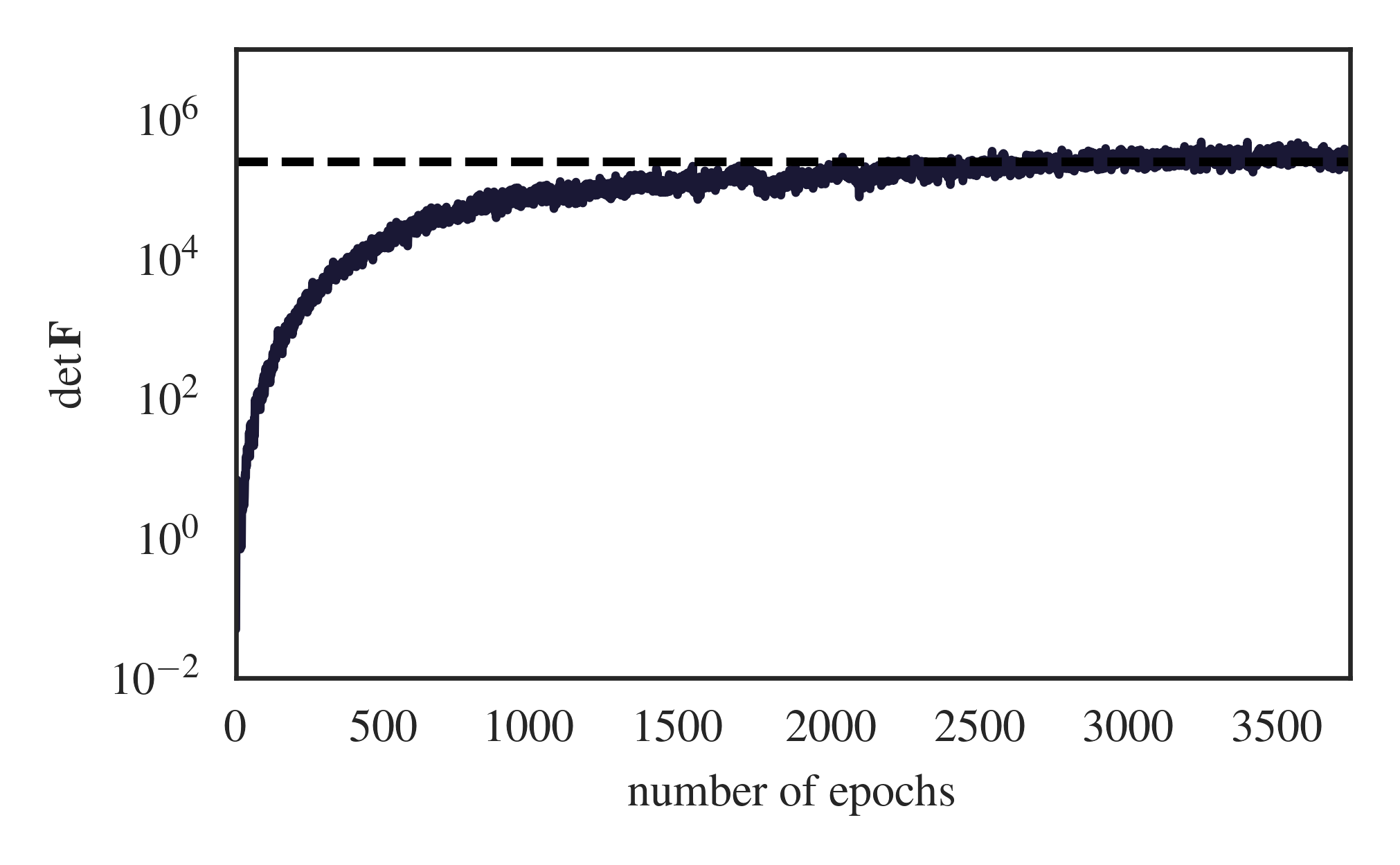}
    \caption{On-the-fly IMNN training using $128\times128$ cosmological field simulations at the initial fiducial model $(\Omega_c, \sigma_8)=(0.6,0.6)$.  Within 1600 epochs of on-the-fly sets of 200 simulations, the network is able to extract $90\%$ of the theoretical Fisher information (dashed black line), saturating to $100\%$ by 3000 training epochs.}
    \label{fig:cosmo-training}
\end{figure}

\begin{figure*}[htp!]
    \centering
    \includegraphics[width=\textwidth]{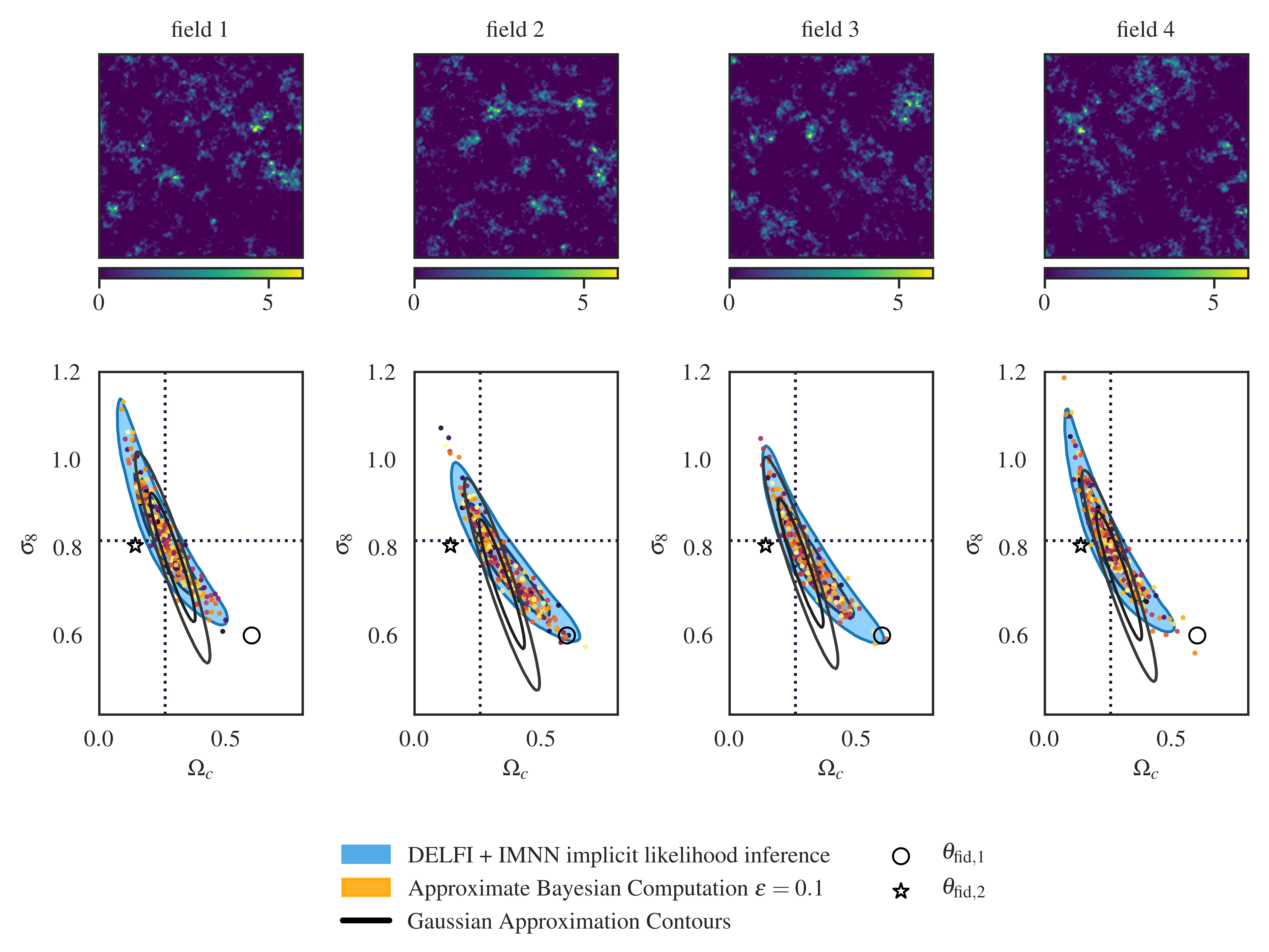}
    \caption{Field-based inference comparison for four independent log-normal cosmological fields (top row). Using IMNN compression immediately gives estimates and Gaussian Approximation uncertainties (black contours) for cosmological parameters, even for a poorly-chosen fiducial model. Even a single iteration of IMNN compression (circle and star) gives summaries that are nearly maximally informative compared to theoretical estimates. Using IMNN summaries, both ABC (orange) and DELFI (blue) methods give consistent posterior contours. DELFI matches ABC compression but requires $\gtrsim$100 times fewer simulations. }
    \label{fig:cosmo-inference}
\end{figure*}

We display target log-normal dark matter fields and inference results in figure \ref{fig:cosmo-inference}. We first train the IMNN on the poor fiducial cosmological model $\parvec_{\rm fid} = (\Omega_c, \sigma_8) = (0.6, 0.6)$, with all other power spectrum parameters consistent with Planck15. Training saturated at $100\%$ of the Shannon information content of the field at the fiducial model, shown in figure \ref{fig:cosmo-training}. We then performed a Gaussian Approximation on the four target fields using the IMNN's Fisher matrix. Since the IMNN's parameter estimates move more than 1$\sigma$ from the fiducial model, we train a new IMNN at the score estimates, $\parvec_{\rm fid, 2} = (0.229, 0.745)$, saturating to 98\% of the corresponding Shannon information. With the compression closer to the target, we then train a DELFI ensemble for each target data over the wide priors $\mathcal{P}(\Omega_c) = \mathcal{U}(0,1.5)$ and $\mathcal{P}(\sigma_8) = \mathcal{U}(0, 1.5)$ for 4 iterations, recovering the blue posteriors needing a total of only 4000 simulations per target data inference. We also compare the DELFI estimates to more expensive ABC sampling, for which we ran 50,000 simulations per target datum with an acceptance parameter $\epsilon=0.1$. Accepted points ($\approx 250$ per field) are coloured according to the log of the distance calculation in equation \ref{eq:abc-distance}.

\section{Discussion \& Conclusion}

In this study, we compared posteriors obtained from implicit likelihood inference via optimally compressed nonlinear summaries to exact likelihood computation for Gaussian field inference. We show that nonlinear summaries from sufficiently expressive convolutional neural networks saturate the known Fisher information content in Gaussian fields. Implicit likelihood inference with this optimal compression yields the exact analytic posterior obtained via the full-field analytic likelihood computation, meeting the benchmark for implicit likelihood inference. We then apply the same network to log-normal fields generated by a more realistic cosmological power spectrum, demonstrating the applicability of our pipeline.

We also introduced new aspects to the IMNN framework. Implementing the network and simulator in \texttt{Jax}'s fully-differentiable framework meant having exact derivatives of simulation parameters with respect to the compression network, eliminating the need for validation and finite difference derivative datasets, significantly reducing the amount of memory needed for network training. 

We also demonstrated IMNN training in an iterative fashion where the initial fiducial model is poorly chosen. Since the IMNN compression is trained at a parameter point (rather than using parameters drawn from a prior) a network trained far from the target data cannot be expected to be maximally informative about the target, leading to suboptimal inference. To address this, we performed a single iteration of  IMNN training for four mock dark matter log-normal fields. Although initially trained far from the target, the target data parameter estimates obtained from the first IMNN allowed the new compression to be trained much closer to the posterior mean, yielding tighter posteriors with the updated IMNN.

The results of this work hold several implications for cosmological parameter estimation. With a Fisher information-maximizing nonlinear compression, the full information of a cosmological field can be represented by optimal summaries, allowing for likelihood-free or simulation-based inference, resulting in posteriors for cosmological or astrophysical parameters. The equivalence of these posteriors based on  IMNN summaries to the analytic solution for Gaussian fields demonstrates that given a sufficiently expressive convolutional network compression, implicit likelihood inference with nonlinear summaries can yield near-exact posteriors. 

\cite{leclercq2021accuracy} raised the question whether implicit likelihood, or simulation-based, approaches can give results that are comparable to full field-based implementations of Bayesian Hierarchical Models, especially for non-Gaussian fields with heavy tail probabilities, such as log-normal data. While we do not directly study the specific example used in that work, our results seem to provide evidence that IMNNs provide near-optimal summaries for  correlated log-normal fields with cosmological power spectra. It can therefore be hoped that the massive compression techniques we describe here remain promising for cosmological parameter estimation as a complement to bespoke sampling implementations that  marginalize high-dimensional latent parameters in  Bayesian Hierarchical Models with complex physical modeling and observational effects \citep[see e.g.][]{jasche_bayesian_2013,jasche2015,lavaux_unmasking_2016,Jasche2019_PM,pydelfiAlsing_2019}. 

Follow-up study is warranted on much larger (e.g. three-dimensional) datasets and with more parameters,  with a view to readying this framework for application to inference from detailed physical simulations and, ultimately, realistic datasets.

\section{Code Availability}
The code used for this analysis is available at \url{https://github.com/tlmakinen/FieldIMNNs}. Full documentation for the IMNN software is available at \url{https://www.aquila-consortium.org/doc/imnn/index.html}.

\begin{acknowledgments} 
T.L.M acknowledges the hospitality of SISSA, Trieste, where part of this work was completed; and Master's scholarship funding from Sorbonne University.
B.D.W. acknowledges support by the ANR BIG4 project, grant ANR-16-CE23-0002 of the French Agence Nationale de la Recherche;  and  the Labex ILP (reference ANR-10-LABX-63) part of the Idex SUPER, and received financial state aid managed by the Agence Nationale de la Recherche, as part of the programme Investissements d'avenir under the reference ANR-11-IDEX-0004-02.
The Flatiron Institute is supported by the Simons Foundation. T.C. acknowledges financial support from the Sorbonne Univ. Emergence fund, 2019-2020. J.A. was supported by the research project grant Fundamental Physics from Cosmological Surveys funded by the Swedish Research Council (VR) under Dnr 2017-04212.
\end{acknowledgments}

\bibliography{mybib}% Produces the bibliography via BibTeX.
\bibliographystyle{aasjournal}

\myemptypage
\newpage
\appendix 

\section{The Fisher Matrix For a Gaussian Random Field With Parameter-Dependent Covariance}\label{app:fisher-field-comp}

Working from equation \ref{eq:field-like}, we can arrive at the Fisher matrix by Taylor expanding the log-likelihood about a set of fiducial parameters $\theta^{(0)}$:
\begin{equation}\label{eq:approximation}
    \hat{\theta} \approx \theta^{(0)} - \frac{\ln \like_{,\theta}(\theta^{(0)})}{\ln \like_{, \theta \theta}(\theta^{(0)})}
\end{equation}

With our likelihood defined in Eq. \ref{eq:field-like}, we can proceed to compute the derivatives needed for our maximum likelihood estimate:
\begin{align}\label{eq:first-deriv}
    \frac{\partial}{\partial \theta} \ln \like &= \frac{\partial}{\partial \theta} \left( -
    \frac{1}{2} \ln (\det \Covar) - \frac{1}{2} \deltdat^T \Cinv \deltdat \right) \\
    &= \frac{\partial}{\partial \theta} \left( -
    \frac{1}{2} \tr (\ln \Covar) - \frac{1}{2} \deltdat^T \Cinv \deltdat \right) \\
    &=   -
    \frac{1}{2} \tr (\Cinv \Covar_{, \theta}) - \frac{1}{2} (\deltdat^T \Cinv \Covar_{, \theta} \Cinv \deltdat ) \\
\end{align}
where we've used the fact that $\ln( \det \Covar) = \tr(\ln \Covar)$ and $\Cinv_{, \theta} = -\Cinv \Covar_{, \theta} \Cinv$. Proceeding to the second derivative,
\begin{multline}\label{eq:second-deriv}
    \frac{\partial^2}{\partial \theta^2} \ln \like = -\deltdat^T \Cinv \Covar_{, \theta} \Cinv \Covar_{, \theta} \Cinv \deltdat \\
    + \frac{1}{2} \tr ( \Cinv \Covar_{, \theta} \Cinv \Covar_{, \theta} ) \\
    + \frac{1}{2} ( \deltdat^T \Cinv \Covar_{, \theta \theta} \Cinv \deltdat - \tr (\Cinv \Covar_{, \theta \theta}))
\end{multline}
The second derivative describes the \textit{curvature} of the likelihood surface, usually notated $\mathcal{F} = -\frac{\partial^2 \like}{\partial \theta^2}$. Taking the expectation $\deltdat \deltdat^T \rightarrow \langle \deltdat \deltdat^T \rangle = C$ in the second derivative, the last term of Eq. \ref{eq:second-deriv} vanishes, allowing us to write Eq. \ref{eq:approximation} as 
\begin{equation}\label{eq:fisher-base}
    \hat{\theta} \approx \theta^{(0)} + \fisher^{-1}_{\theta \theta} \frac{\deltdat^T \Cinv \Covar_{, \theta} \Cinv \deltdat - \tr(\Cinv \Covar_{, \theta})}{2}
\end{equation}
where $\textbf{F}_{\theta \theta}$ is the Fisher matrix. This can be generalized to a multi-parameter case where the estimator for a parameter is given as:
\begin{equation}
    \hat{\theta}_\alpha \approx \theta^{(0)}_\alpha + \fisher^{-1}_{\alpha \beta} \frac{\deltdat^T \Cinv \Covar_{, \beta} \Cinv \deltdat - \tr(\Cinv \Covar_{, \beta})}{2}
\end{equation}
where the Fisher matrix for parameters indexed by $\alpha, \beta$ is defined as
\begin{equation}\label{eq:fisher1}
    \fisher_{\alpha \beta} = \langle \mathcal{F} \rangle= \frac{1}{2} \tr (\Covar_{, \alpha} \Cinv \Covar_{, \beta} \Cinv).
\end{equation}

\section{Generating Gaussian Fields in Jax}\label{sec:makesims}
To generate simulated Gaussian fields numerically for
testing our likelihood-free inference pipeline, we adopt
the following recipe:

\begin{enumerate}
    \item Generate a unit normal white noise field in
    $k$-space, $\varphi_\textbf{k}$ such that $\langle
    \varphi_\textbf{k} \varphi_{-\textbf{k}}\rangle = 1$
    \item Satisfy the reality condition
    \item Scale white-noise field by the square-root of the
    power spectrum: $R_{\rm P} = P^{1/2}(k) R_{\rm
    white}(\textbf{k})$
    \item Fourier transform the scaled field back to real
    space, $R_{\rm P}(\textbf{x}) = \int d^d \tilde{k} e^{i
    \textbf{k}\cdot \textbf{x}} R_{\rm P}(\textbf{x})$
\end{enumerate}

 We begin by generating the white noise field in $k$-space,
 taking care to ensure that the resulting field is
 Hermitian. Moving from a continuous description to a
 discrete one, a 2D field is now represented by an $N\times
 N$ grid $\phi^\textbf{x}_{ab}$ where the indexes $a,b \in
 \{0, \dots, N-1 \}$. This means that the Fourier transform
 that relates $\phi(\textbf{x})$ to its conjugate
 $\phi(\textbf{k})$
 is now discrete, reading (in Numpy notation):
 \begin{align}
     \phi_{ab}^{\textbf{k}} &= \sum_{c,d = 0}^{N-1}
     \exp{(-i x_c k_a - i x_d k_b)
     \phi^{\textbf{x}}_{cd}} \\
    \phi_{ab}^{\textbf{x}} &= \frac{1}{N^2}\sum_{c,d =
    0}^{N-1} \exp{(-i x_c k_a - i x_d k_b)
    \phi^{\textbf{k}}_{cd}}
\end{align}
When generating random fields, it is important that Hermitianity (the reality condition) is ensured. This means that given a real $\phi^\textbf{x}_{ab}$ and even number of grid points $N$, the conjugate field satisfies (for an integer $\alpha$):
\begin{align}
    \phi^{\textbf{k}}_{ab} &= \phi^\textbf{k}_{(\alpha +\alpha N)b} = \phi^\textbf{k}_{a(b +\alpha N) } \\
    \phi^{*\textbf{k}}_{ab} &= \phi^\textbf{k}_{-a,-b}
\end{align}
this also implies that $\phi^{*\textbf{k}}_{(a(N/2 + b))} = \phi^\textbf{k}_{a(N/2 - b)}$ and $\phi^{*\textbf{k}}_{(b(N/2 + a))} = \phi^\textbf{k}_{b(N/2 - a)}$. To do this numerically, we draw the magnitude, $m$ of each pixel of the random field $\phi(\textbf{k})$ at random from a unit variance normal distribution,
\begin{align}
    m^\textbf{k}_{ab} &\sim \mathcal{N}(0, 1) \\
    m^\textbf{k}_{ab} &\gets \frac{m_{ab} + m_{ba}}{\sqrt{2}}
\end{align}
where to satisfy Hermitianity the magnitude array with reversed indexes is summed to the original $m^\textbf{k}_{ab}$
using the same random key, we draw the complex phases from a uniform distribution between $0$ and $2 \pi$:
\begin{align}
    p^\textbf{k}_{ab} &\sim \mathcal{U}(0, 2\pi ) \\
    p^\textbf{k}_{ab} &\gets \frac{p_{ab} + p_{ba}}{2} + \pi
\end{align}
The complex white noise field is then obtained via Euler's formula 
\begin{equation}
    \phi^\textbf{k}_{ab} \gets m^\textbf{k}_{ab} [ \cos(p^\textbf{k}_{ab}) + i \sin(p^\textbf{k}_{ab}) ]
\end{equation}
To satisfy the reality condition for the power spectrum scaling, the $a,b$ index is chosen to be $k_a = k_b = \frac{2 \pi}{N} \{0, \dots N/2, -N/2 + 1, \dots -1 \} \in \mathbb{R}^N$. To evaluate the scaling field, the power spectrum is evaluated for all points on the grid formed by the outer product of the $k_a$ and $k_b$ arrays: 
\begin{align}
    P^\textbf{k}_{ab} &\gets P({k}_a \otimes {k}_b) \\
    R^\textbf{k}_{ab} &\gets P^\textbf{k}_{ab} \phi_{ab}^\textbf{k}
\end{align}
The scaled random field can then be Fourier transformed back to the real-space representation, (with an appropriate correction of $N^2$ to compensate the built-in Numpy implementation). 

We implement these steps using the numerically-differentiable \texttt{Jax} Numpy backend in Python. Since \texttt{Jax} is by default XLA-compilable, care must be taken when assigning array elements to memory, since typical Boolean masking is not enabled for gradient computation. For instance, for $\textbf{k}_{ab}$ index values of 0, a power spectrum $P (k) = Ak^{-B}$ yields an undefined value. The workaround in \texttt{Jax} is to assign values via a preprogrammed conditional operator that selects for this zero mode. 

\subsection{Log-Normal Fields}\label{sec:logfromgauss}
To generate scaled log-normal fields from a specified power spectrum, a few steps are modified from the Gaussian case. First the grid of magnitudes is transformed as
\begin{equation}
    P^\textbf{k}_{ab} \gets \ln ( 1 + P^\textbf{k}_{ab} ) \\
\end{equation}
and multiplied by the gaussian noise. The field is then rescaled by the specified volume of the simulation. Once the real-space field is obtained via the inverse Fourier transform, it is transformed and rescaled by the variance as:
\begin{equation}
    \Phi \gets \exp{\left( \Phi - \frac{\langle \Phi \Phi \rangle}{2} \right) } - 1
\end{equation}

\end{document}